%% file: ms.tex
\documentclass[11pt,a4paper]{article}
\pdfoutput=1
\PassOptionsToPackage{obeyspaces}{url}
\usepackage[mathlines]{lineno}
\usepackage[shortlabels]{enumitem}
\usepackage{pgffor}
\usepackage{graphicx}
\usepackage[font=small,labelfont=bf]{caption}
\usepackage{amssymb,amsmath}
\usepackage{fancyvrb}
\usepackage[normalem]{ulem}
\usepackage{array}
\usepackage{ccicons}
\usepackage{listings}
\usepackage{seqsplit}
\usepackage{multirow}
\usepackage{hepnames}
\usepackage{relsize}
\usepackage{siunitx}
\RequirePackage[mathlines]{lineno}
\RequirePackage[british]{babel}
\RequirePackage{booktabs}

\usepackage{jinstpub}
\newcommand{\corry}{Corryvreckan\xspace}
\newcommand{\apsq}{Allpix\textsuperscript{2}\xspace}

\date{\today}

\abstract{
Significant progress has been made to develop silicon pixel technologies for use in the vertex and tracker regions of the proposed Compact Linear Collider (CLIC) detector design. 
The electron-positron collisions generated by this linear accelerator provide a clean, low-radiation environment for the inner detectors. 
However, physics-driven performance targets, the CLIC beam structure, and occupancies from beam-induced backgrounds place challenging requirements on detector technologies for this region. 
A pixel pitch down to 25$\times$\SI{25}{\square\micro\meter}, material budget $\leq$ 0.2-2\%$X_0$ per layer, average power dissipation of down to \SI{50}{\milli\watt\per\square\centi\meter}, position resolution of 3-\SI{7}{\micro\meter}, and timing resolution as low as \SI{5}{\nano\second} are called for in the vertex and tracking detectors. 
To this aim, a comprehensive R\&D programme is ongoing to design and test silicon pixel detectors to fulfil these specifications, including both monolithic and hybrid devices. 
These studies involve \apsq Monte Carlo and TCAD simulations, advanced \SI{65}{\nano\meter} ASIC and sensor design, laboratory testing, and beam tests of individual modules to determine the required performance parameters. 
The characterisation and simulation modelling of these devices has also lead to the development of a set of tools and software within the CLIC detector and physics (CLICdp) collaboration. 
This publication will present recent results from the technologies being developed and tested in view of the CLIC vertex and tracking detector requirements, such as various monolithic CMOS sensors, and fine pitch hybrid assemblies with planar sensors. 
}

\title{R\&D for the CLIC Vertex and Tracking detectors} 
\author[a,b]{Morag Williams}
\affiliation[a]{University of Glasgow,\\Glasgow, U.K.}
\affiliation[b]{CERN,\\Geneva, Switzerland}
\collaboration[c]{on behalf of the CLICdp collaboration}
\emailAdd{morag.williams@cern.ch}


\graphicspath{{figures/}}
\notoc
\begin{document}
\titlepage
\maketitle
\flushbottom
\input{chapters/introduction}
\input{chapters/requirements}
\input{chapters/hybrid}
\input{chapters/monolithic}

\input{chapters/toolsdeveloped}
\input{chapters/conclusion}
\input{chapters/acknowledgements}
\bibliographystyle{JHEP.bst}
\bibliography{references.bib}

\end{document}

%% file: chapters/introduction.tex
\section{Introduction}
\label{ch:introduction}

The Compact LInear Collider (CLIC) is a proposal for a future high-luminosity, high-energy linear lepton collider.
The accelerator will achieve three different centre-of-mass energies due to its staged design: \SI{380}{GeV}, \SI{1.5}{TeV}, and \SI{3}{TeV}~\cite{CLIC-summary-report-2018}.
The generated electron-positron collisions will provide a clean, low-radiation environment for the inner detectors.
However, physics-driven performance targets, the CLIC beam structure, and occupancies from beam-induced backgrounds place challenging requirements on detector technologies for this region~\cite{detector_tech_report}.

Significant progress has been made to develop silicon pixel technologies for use in the vertex and tracker regions of the proposed CLIC detector design~\cite{CLICdet} via a comprehensive R\&D programme.
These studies have also lead to the development of a set of tools and software within the CLIC detector and physics (CLICdp) collaboration~\cite{detector_tech_report}. 

In Section~\ref{ch:requirements} of this publication, the requirements for the inner detectors at CLIC are discussed.
The recent developements in technologies for the CLIC vertex and tracking detectors are presented in Sections~\ref{ch:hybrid} and~\ref{ch:monolithic} respectively.
These include fine-pitch hybrid assemblies for the vertex detector, and High-Voltage (HV) and High-Resistivity (HR) CMOS technologies for the tracking detector.
The software and tools developed during these studies are highlighted in Section~\ref{ch:toolsdeveloped}.

%% file: chapters/requirements.tex
\section{Requirements for the CLIC inner detectors}
\label{ch:requirements}

The foreseen physics programme at CLIC imposes strict requirements on the design of the vertex and tracking detectors~\cite{detector_tech_report}.
A track-momentum resolution of $\sigma_{P_T} / p_T^2 \leq 2 \times 10^{-5}$ GeV$^{-1}$ is needed in the central detector region for high-momentum tracks.
To accurately reconstruct and enable flavour tagging with clean b-, c-, and light-quark jet separation, a transverse impact-parameter resolution of $\sigma_{d_0}^2 = (\SI{5}{\um})^2$ is required.
In addition, the nanometre transverse bunch sizes of the CLIC beam causes large numbers of Beamstrahlung photons to be created during bunch crossing. 
These photons produce background particles that interact with the sensitive material of the vertex and tracker detector, thus fast time-tagging is needed for pile-up rejection.
A combination of air and liquid cooling systems will be used for the CLIC vertex and tracking detectors, therefore the use of a power pulsing scheme is envisaged for a low average power dissipation in the inner detectors~\cite{powerpusling_tpx3}~\cite{powerpusling_cpx2}.

The combination of these leads to the technology requirements listed in Table~\ref{tab:requirements}.
Currently no available detector technologies can fulfil all of these requirements simultaneously.
Consequently, novel sensors and readout technologies are being developed and tested within the collaboration.

\begin{table}[tbp]
\caption{Requirements for the CLIC vertex and tracking detectors~\cite{detector_tech_report}.\label{tab:requirements}}
\centering
\begin{tabular}{lll}
\toprule
\textbf{Quantity}                                    & \textbf{Vertex detector}               					& \textbf{Tracking detector}  \\
 \midrule
\multirow{1}{*}{\textit{Single point resolution}}    & \SI{3}{\micro\meter}							& \SI{7}{\micro\meter}                     \\
  \midrule
\multirow{1}{*}{\textit{Maximum pixel size}}         & $\leq$ \SI{25}{\micro\meter} $\times$ \SI{25}{\micro\meter}		& $\leq$ 30 - \SI{50}{\micro\meter} $\times$ 1 - \SI{10}{\milli\meter}        \\
\midrule
\multirow{1}{*}{\textit{Material budget per layer}}  & 0.2\% $X_0$								& 1 - 2\% $X_0$                  \\
\midrule
\textit{Timing resolution}                           & $~$\SI{5}{\nano\second}							& $~$\SI{5}{\nano\second}                        \\
\midrule
\multirow{1}{*}{\textit{Hit efficiency}}             & 99.7 - 99.9\%                           					& 99.7 - 99.9\%               \\
\midrule
\textit{Average power dissipation}		     & $<\SI{50}{\milli\watt\per\square\centi\meter}$				& $<\SI{150}{\milli\watt\per\square\centi\meter}$\\
\midrule
\textit{Sensor surface}				     & \SI{0.84}{\square\meter}							& \SI{137}{\square\meter}\\
\bottomrule
\end{tabular}
\end{table}

%% file: chapters/hybrid.tex
\section{Hybrid pixel detector technologies for the vertex detector}
\label{ch:hybrid}
\subsection{CLICpix2 with planar sensors}
\begin{figure}[tb]
\centering
\begin{minipage}{.48\textwidth}
    \centering
    \includegraphics[width=\linewidth]{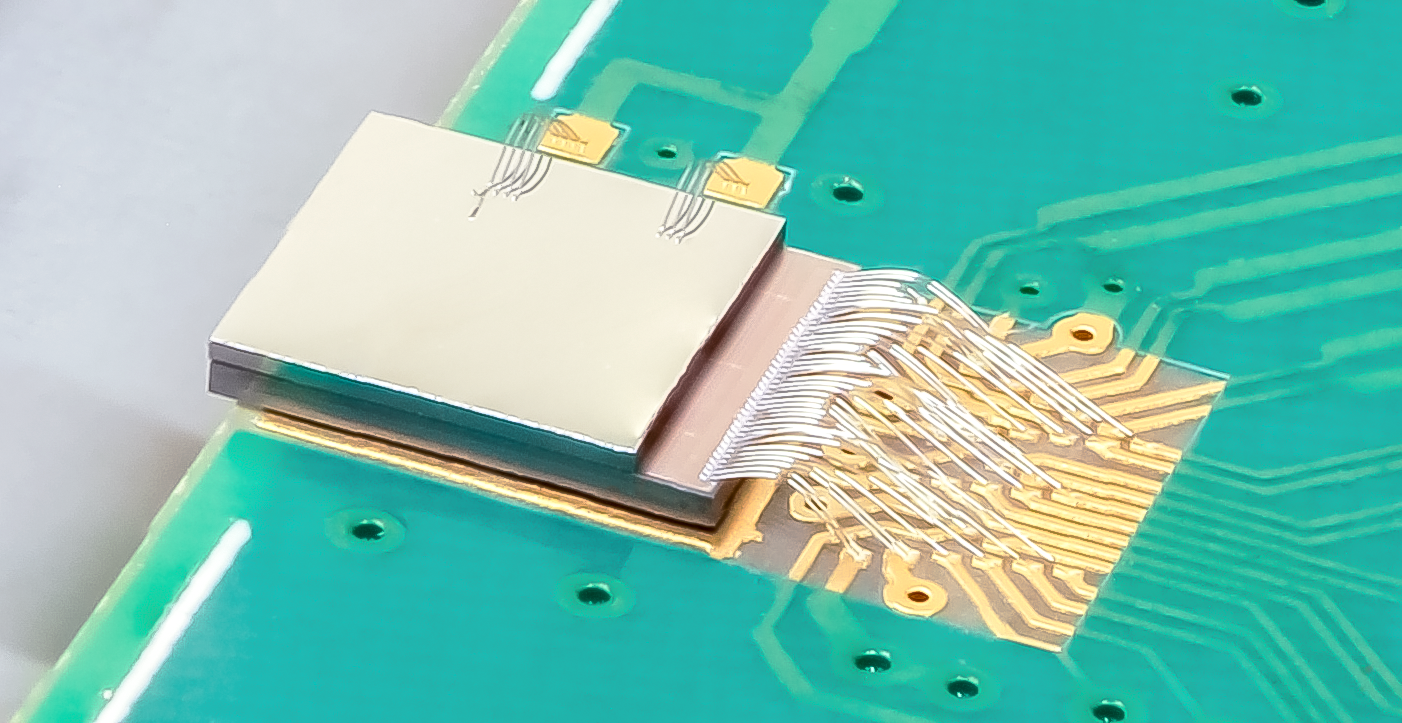}
    \caption{Image of a CLICpix2 ASIC wire-bonded to a readout PCB (bottom)~\cite{detector_tech_report}.}
    \label{fig:CPX2_photo}
\end{minipage}%
\hfill
\begin{minipage}{.48\textwidth}
    \centering
    \includegraphics[width=\linewidth]{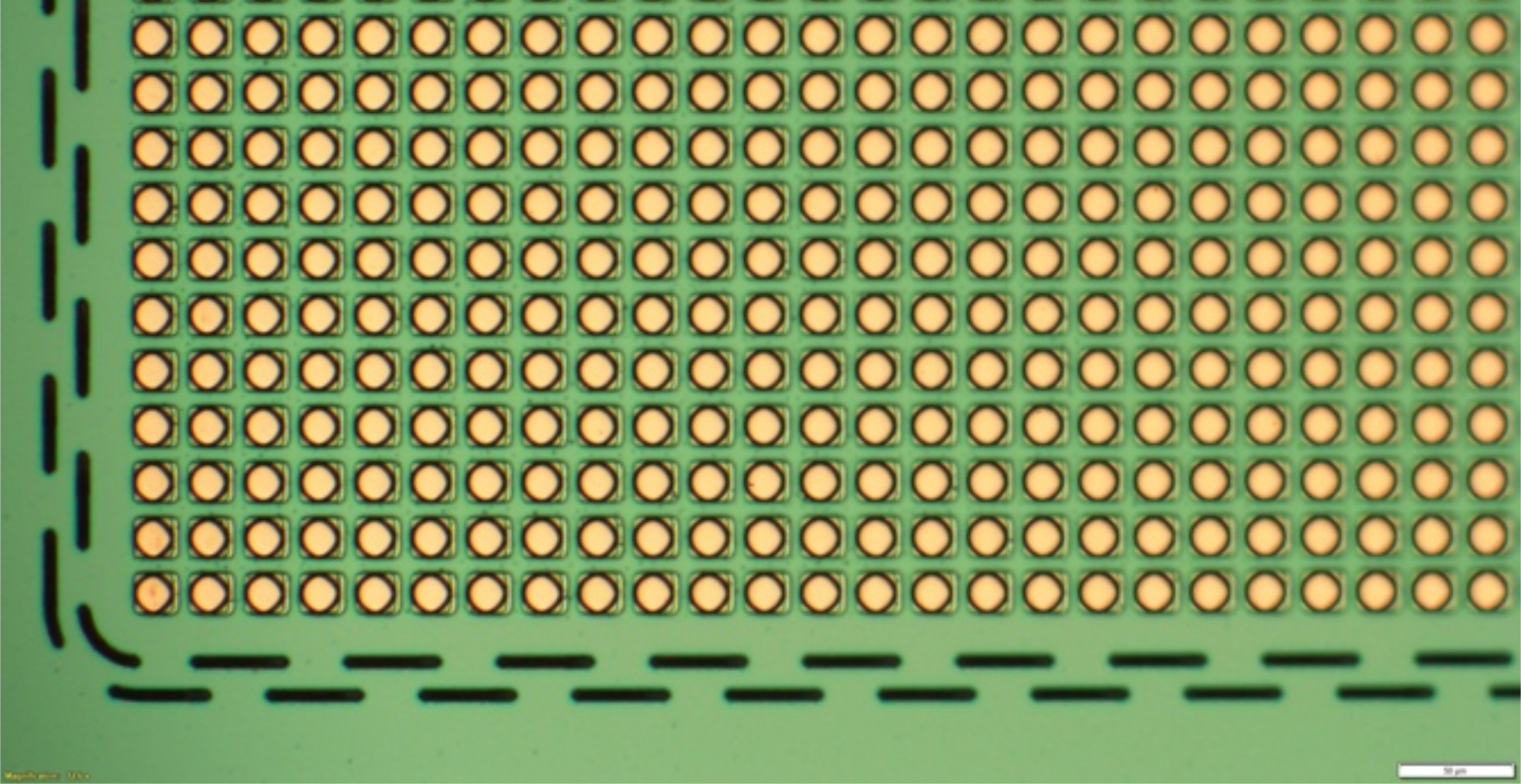}
    \caption{Picture of an FBK CLICpix2 sensor showing the staggered active-edge design.}
    \label{fig:CPX2_staggeredactiveedge}
\end{minipage}  
\end{figure}

CLICpix2, see Figure~\ref{fig:CPX2_photo}, is a small-pitch, pixelated readout chip produced in a commercial \SI{65}{\nano\meter} CMOS process~\cite{cpx2_manual}.
It is part of the Timepix/Medipix family and is the successor to the CLICpix chip.
CLICpix2 ASICs have been solder bump-bonded to planar silicon sensors of thicknesses \SI{50}{\um} to \SI{200}{\um}~\cite{detector_tech_report}.
Figure~\ref{fig:CPX2_staggeredactiveedge} shows an image of a \SI{130}{\micro\meter} thick active-edge sensor.

\textbf{Detector design :} The CLICpix2 pixel matrix consists of 128$\times$128 square pixels of pitch $\SI{25}{\um}$~\cite{cpx2_manual}.
A frame-based readout scheme is employed, where the in-pixel digital back-end can record three different measurements: Time-over-Threshold (ToT), Time-of-Arrival (ToA), and event counts.
These are combined into four CLICpix2 operational modes: simultaneous 8-bit ToA and 5-bit ToT, simultaneous 8-bit count and 5-bit ToT, 13-bit ToA, or 13-bit measurement of counts~\cite{cpx2_manual}. 

\textbf{Laboratory results :} A single-chip bump-bonding process is being developed at IZM~\cite{IZM} for CLICpix2 ASICs and active-edge sensors from FBK-CMM~\cite{FBK} and Advacam~\cite{Advacam}.
Each of these devices were tested in the laboratory using a combination of threshold equalisation, I-V scans, test pulses, and radioactive source exposures.

\textit{Test pulse measurements:} Charge of a known magnitude was directly injected into the CLICpix2 ASIC frontend so that the ASIC response per pixel could be seen independently from the sensor and bonding quality.
In the data analysis, pixels that recorded an event fewer than 10\% of the total number of pulses sent are said to be 'unresponsive'. 
This is either due to the pixel readout being masked due to high noise, being electrically connected to a large number of shorted pixels, or the pixel has a non-functioning frontend. 
Some pixels were found to record hits while other pixels were test pulsed, and if this occurs during $\geq$90\% of the test pulses of a neighbouring pixel then these are called 'unexpected' pixel responses.
This unwanted response can be caused by electrical connections between neighbouring pixels or cross-talk.

\textit{Radioactive source exposure:} To investigate the behaviour of each CLICpix2 assembly to ionising particles, a Sr-90 radioactive source was placed above the assembly to obtain evenly distributed hits across the pixel matrix. 
A histogram of the pixel hit rates is then fitted by a Gaussian function and pixels are expected to be within 3$\sigma$ of the Gaussian mean.
The tails above and below the 3$\sigma$ cut correspond to pixels that are deemed to be 'high rate' or 'low rate' respectively.
The Gaussian peak contains 99.7\% of all pixels, therefore in a perfect assembly 25 pixels are expected to be high rate and 25 to be low rate.  

\textit{Pixel categorisation:} To better understand the assembly performance, the results from the described laboratory measurements are combined to classify pixels into six categories. 
These data-driven category definitions are detailed below.
It is noted that each pixel can only be in one category, with preceding categories taking precedence. 
\begin{enumerate}
    \item Masked: Pixels of this category have been masked during threshold equalisation or by the user due to large noise. No information is gathered from these pixels, therefore their response cannot be characterised.
    \item Unresponsive: Pixels in this category responded less than 10\% of the time to charge injection.
    \item Shorted: The pixel records unexpected hits when neighbouring pixels are pulsed $\geq$90\% of the time. 
    \item Bonding or sensor issues: Pixels in this set were found to be low rate but have the expected test pulse performance, so are said to have bonding or sensor issues. 
    \item High rate: This set of pixels were found to have a high rate during a source measurement. 
    \item Expected response: The expected responses are seen in all tests for this pixel: normal test pulse response, normal rate from source measurement, no other issues.
\end{enumerate}
The categorisation results for assembly 20 are presented in Table~\ref{tab:cpx2_categories} and Figure~\ref{fig:cpx2_categories}.
The majority of high rate pixels for this device are located 1-2 rows/columns from the matrix edge. 
One possible explanation for this effect is that the electric field distribution from the staggered active edge design, see Figure~\ref{fig:CPX2_staggeredactiveedge}, causes rows/columns 1-2 to receive a lower amount of charge, while rows/columns 2-3 receive a higher amount.
Using the categorisation results, an interconnect yield of $(expected\_response+high\_rate)/(128\times128 -masked -unresponsive) = 97.9\%$ was calculated.
This is the highest interconnect yeild obtained for a CLICpix2 planar assembly so far.
The yeild of high-quality assemblies is low due to the challenging bump-bonding process for the small pitch size. 
It was found that the combination of Ti/W/Cu Under Bump Metalisation (UBM) used for the FBK-CMM sensors and SnAg bump material used by IZM produced the best results.

\begin{table}[tb]
  \begin{minipage}[c]{0.4\textwidth}
    \captionof{table}{Summary of categorisation results of CLICpix2 assembly 20.}
    \label{tab:cpx2_categories}
    \centering
    \begin{tabular}{ll}
      \toprule
      \multirow{2}{*}{\textbf{Category}}  & \textbf{Number}\\
                                          & \textbf{of Pixels}              \\
    \midrule	
	  1. Masked			& 32\\
    \midrule
	  2. Unresponsive		& 2\\
      \midrule
	  3. Shorted			& 2\\
    \midrule
	  4. Bonding or			& \multirow{2}{*}{344}\\
	     sensor issues		&		\\
    \midrule
	  5. High rate		   	& 247\\
    \midrule
	  6. Expected response		& 15757\\
    \bottomrule
    \end{tabular}
  \end{minipage}\hfill
  \begin{minipage}[c]{0.55\textwidth} 
    \centering
    \includegraphics[width=\linewidth]{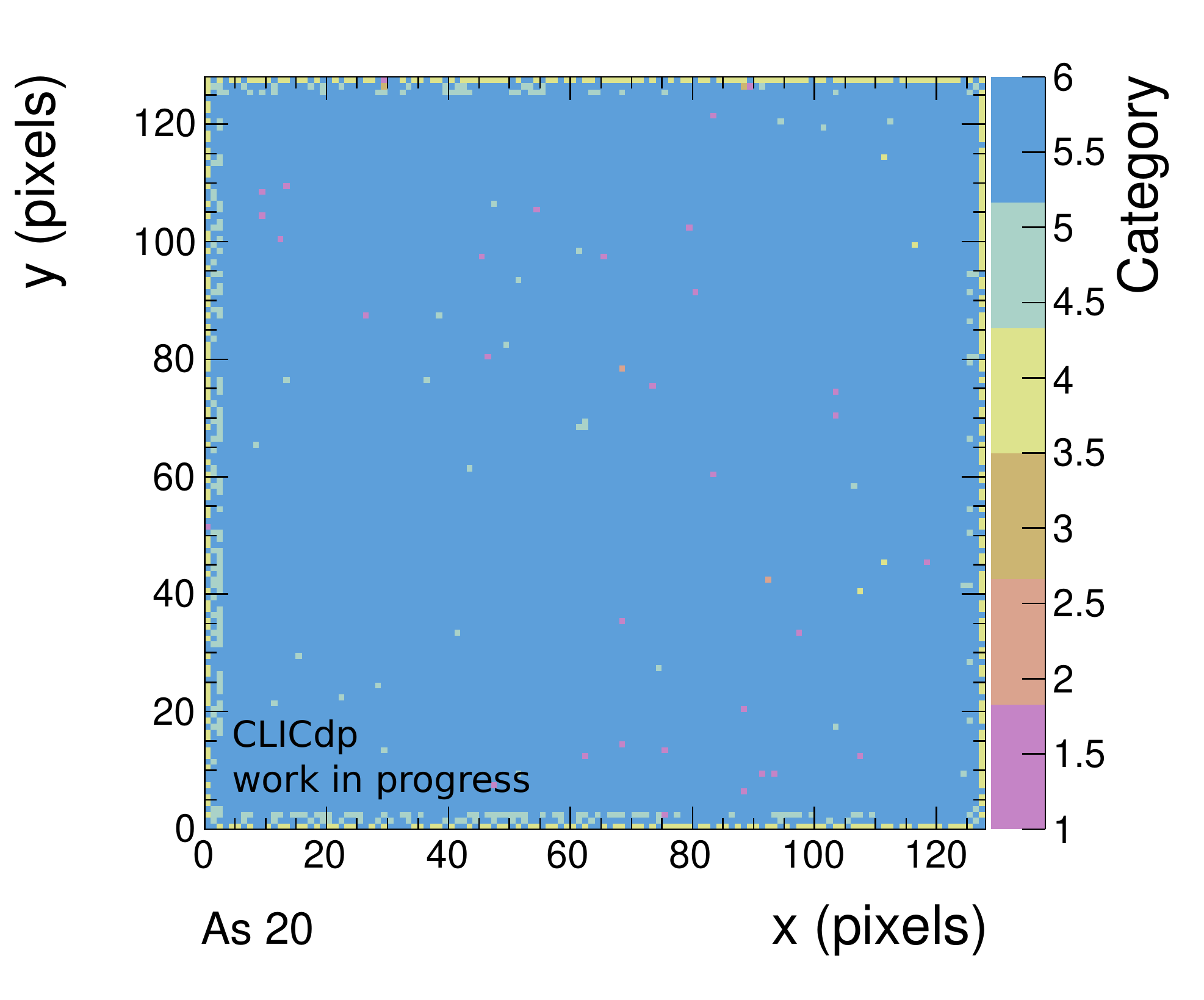}
    \captionof{figure}{A 2D map showing the categories determined for each pixel of CLICpix2 planar assembly 20.}
    \label{fig:cpx2_categories}
  \end{minipage}
\end{table}

\textbf{Test-beam results :} Assembly 20 was tested using the DATURA beam telescope set-up~\cite{EUDET_telescope_paper} in the DESY II test-beam area~\cite{DESYpaper}.
The DESY-II synchrotron supplies an electron/positron particle beam of energies up to \SI{6}{GeV}, and a nominal beam energy of \SI{5.4}{GeV} was chosen for data taking.
For the set-up used, the achievable telescope tracking resolution at the DUT plane was calculated to be \SI{2.5}{\um}.
The nominal operating conditions for CLICpix2 assembly~20 were determined during the laboratory testing: a threshold of approximately \SI{645}{electrons} and a bias voltage of \SI{-60}{V}.

\begin{figure}[tb]
 \begin{minipage}[t]{0.48\textwidth} 
 \centering
    \includegraphics[width=\linewidth]{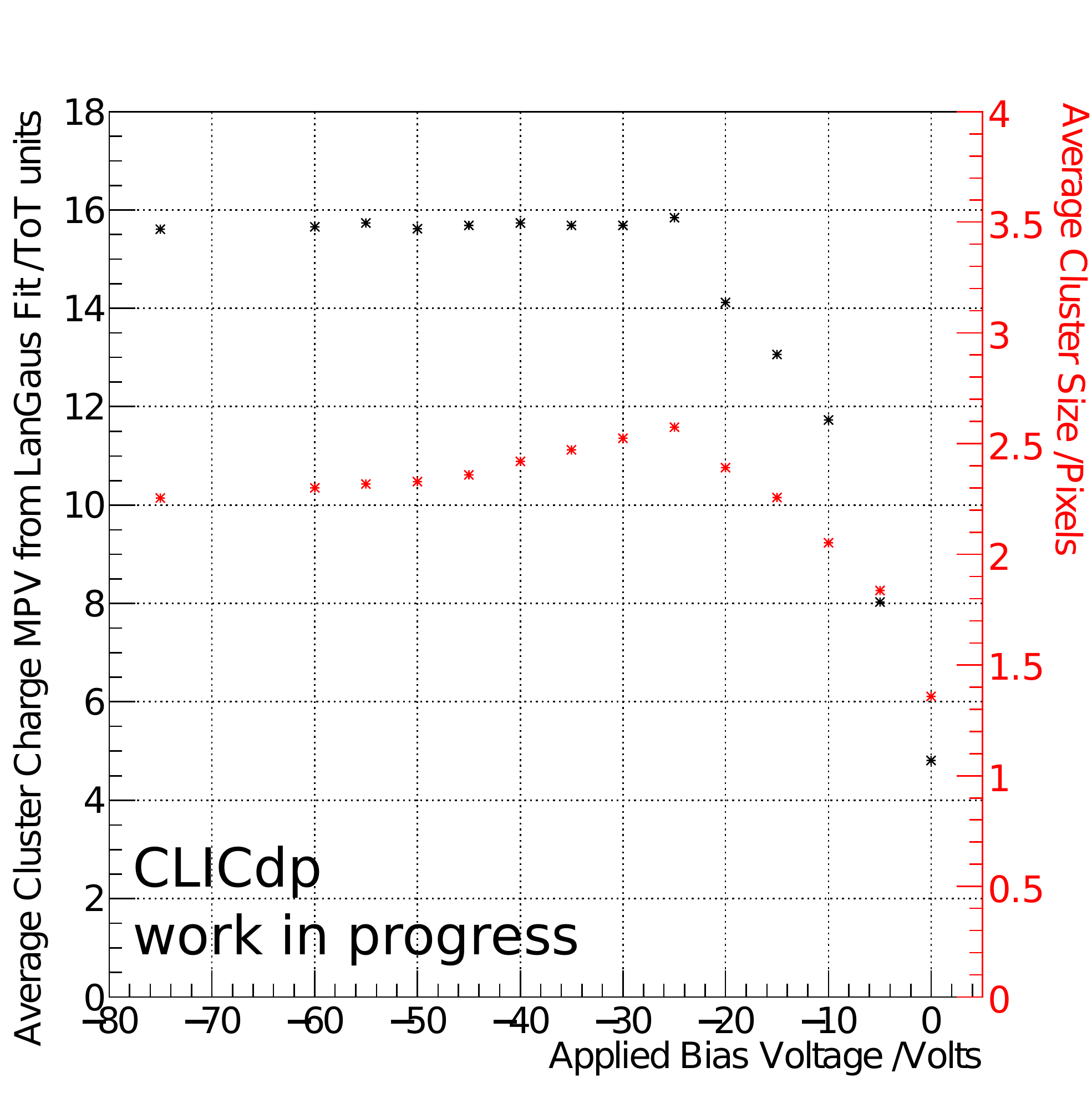}
    \caption{Histogram of mean track-associated cluster size and MPV of cluster charge vs applied bias voltage for CLICpix2 planar assembly 20.}
    \label{fig:cpx2_voltagescan_clusterplots}
 \end{minipage}\hfill
 \begin{minipage}[t]{0.48\textwidth} 
 \centering
    \includegraphics[width=1.05\linewidth]{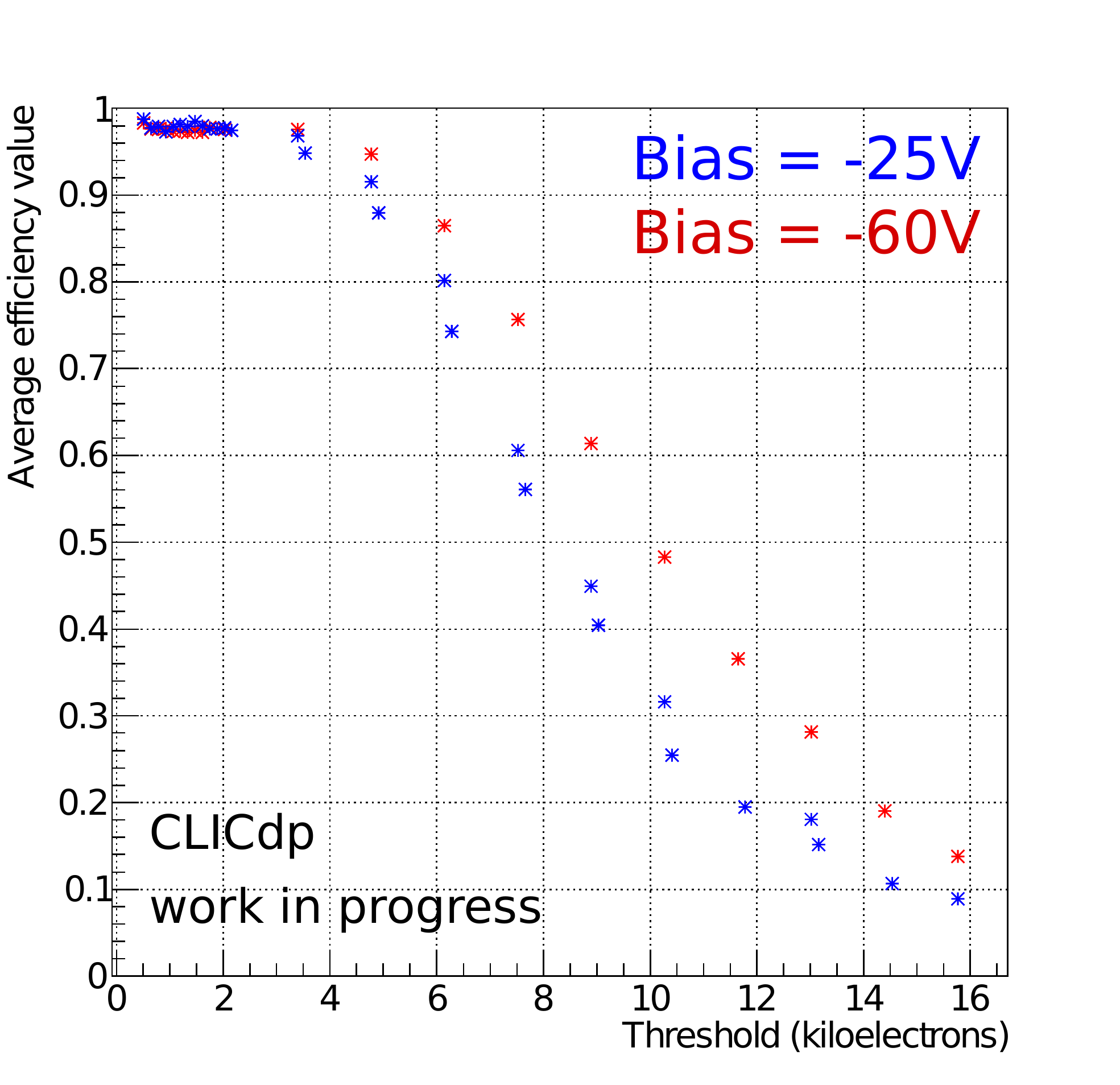}
    \caption{Effiency as a function of threshold for CLICpix2 assembly 20 at bias voltages of \SI{-25}{V} and \SI{-60}{V}.}
    \label{fig:cpx2_thresholdscan_eff}
 \end{minipage}
\end{figure}

By varying the bias voltage applied, the depletion voltage and optimal applied bias voltage can be determined.
Figure~\ref{fig:cpx2_voltagescan_clusterplots} presents such a scan for assembly 20 with two observable quantities: the mean size of track-associated clusters and the Most Probable Value (MPV) from a Landau-Gaussian fit to the associated cluster charge distribution.
The bias voltage that produces the highest mean cluster size in this scan, and therefore the maximum amount of charge sharing, is \SI{-25}{V}.
This is therefore the optimal operational bias voltage for this device.
To calculate the depletion voltage of the assembly, two linear functions were fit to the histogram of MPV$^2$ vs bias voltage.
The intersection of these two functions provides the depletion voltage, which was calculated to be~\SI{-24}{V} for this sample.

Test-beam analysis has calculated the efficiency of this assembly to be 98.1\% at nominal conditions.
The device detection threshold was scanned from \SI{0.5}{ke} to \SI{15.9}{ke}, and the resultant efficiency trends are shown in Figure~\ref{fig:cpx2_thresholdscan_eff}.
The efficiency falls with increasing threshold as expected.

\subsection{ELADs}
The positional resolution of thin sensors is often limited to the binary hit resolution of $pitch/\sqrt{12}$ due to the limited charge sharing when traversing the small sensor thickness.
Reducing the pixel size can improve this, however the pitch is often limited by the feature sizes of the readout electronics.
The Enhanced LAteral Drift (ELAD) design enhaces the lateral drift in the sensor layer to increase charge sharing between adjacent pixels.
ELAD sensors allow the positional resolution to be improved without the need to adapt the pixel pitch and/or sensor thickness~\cite{elad}.

\begin{figure}[tb]
\begin{minipage}[t]{0.24\textwidth} 
    \centering
    \includegraphics[width=0.94\linewidth]{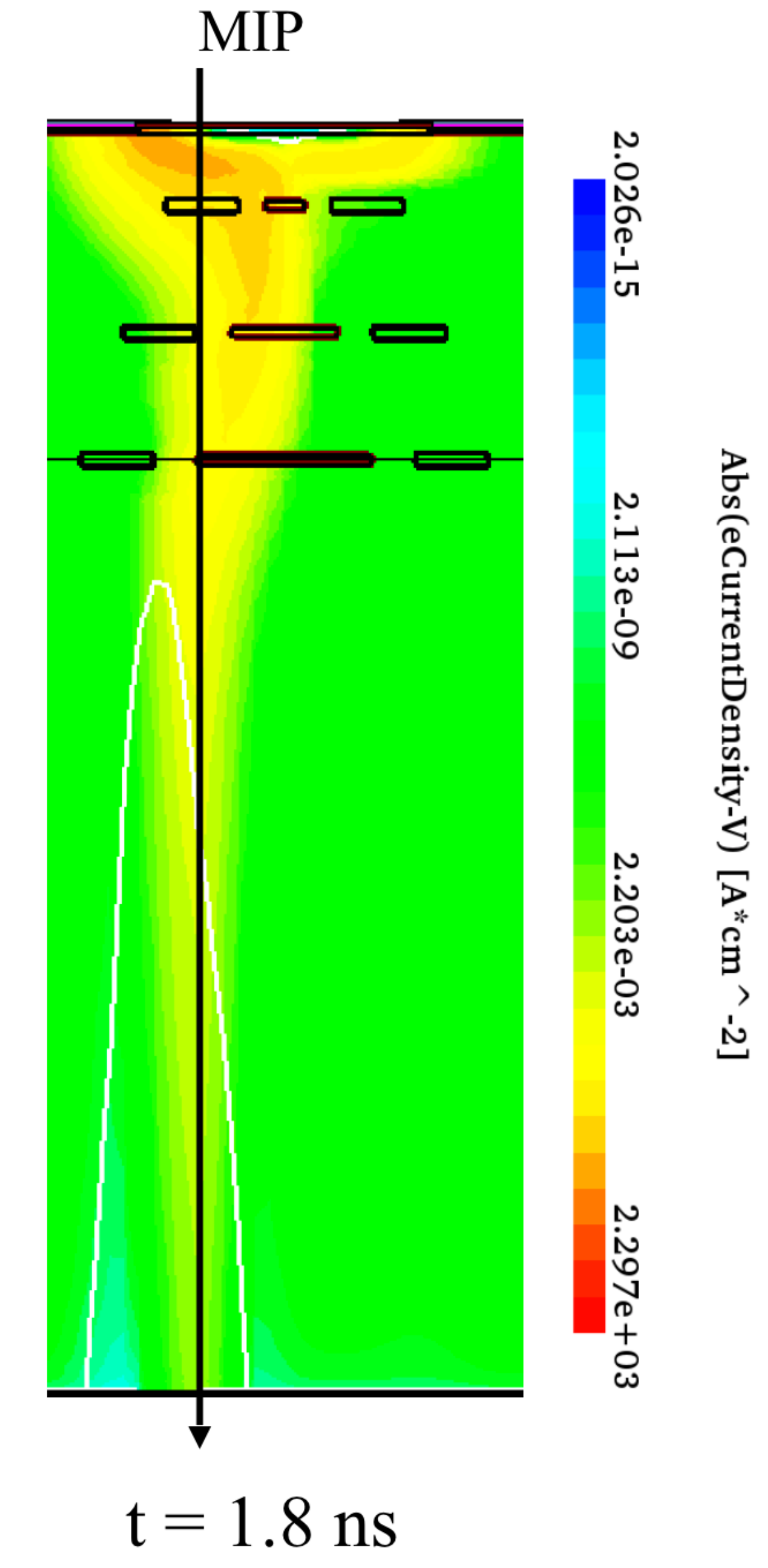}
    \caption{TCAD transient simulation of a \SI{150}{\micro\meter} thick ELAD sensor after an incident MIP. Modified from~\cite{detector_tech_report}.}
    \label{fig:elad_tcad}
 \end{minipage}\hfill
 \begin{minipage}[t]{0.72\textwidth}
    \centering
    \includegraphics[width=\linewidth]{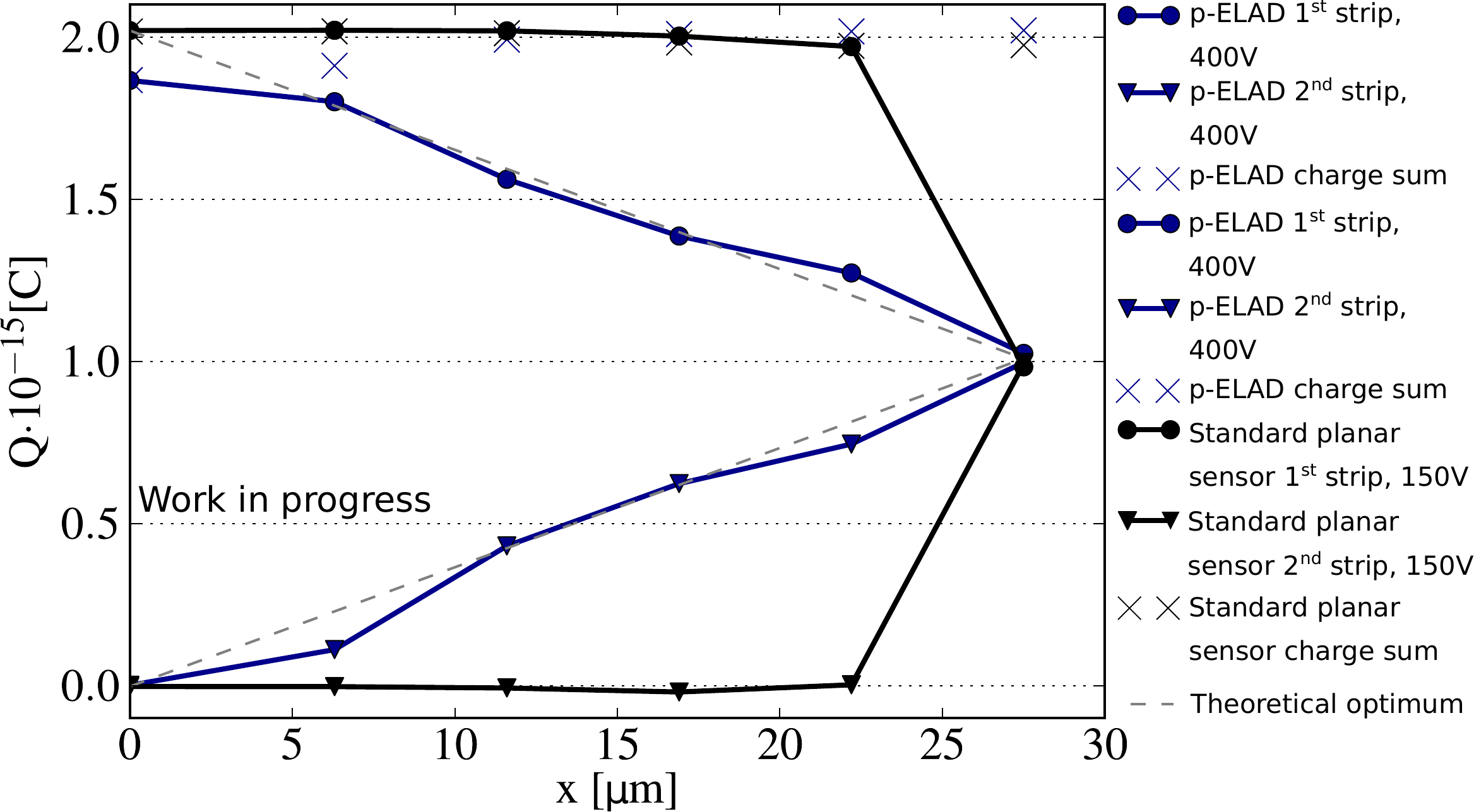}
    \caption{TCAD simulation of the charge collected by adjacent strips in an ELAD sensor vs. distance.
    A position of \SI{0}{\um} corresponds to the center of the first strip, and \SI{27.5}{\um} to the center between the two strips.
    The results are compared to standard sensor biased at \SI{150}{\volt}.
    Modified from~\cite{detector_tech_report}.}
    \label{fig:elad_resolution}
 \end{minipage}
\end{figure}

\textbf{Simulation results :} Deep ion implants inside the sensor are used to adapt the electric field to both promote lateral drift and reduce low electric field regions~\cite{elad}. 
The position, dimensions, number, and doping concentrations of the implants were simulated using 2D and 3D TCAD finite element simulations~\cite{synopsys-tcad} to optimise the sensor design.
A pixel sensor with \SI{55}{\micro\meter} readout electrode pitch was assumed.
In Figure~\ref{fig:elad_tcad} the implant design developed from these investigations can be seen as a cross-section in a TCAD transient simulation.
These implants are simulated as strips, such that charge sharing is enhanced along one lateral axis.
The electron current density is plotted \SI{1.8}{\nano\second} after a Minimally Ionising Particle (MIP) has been simulated to hit the sensor surface at a distance of \SI{16.9}{\um} from the left-hand pixel implant.

The detector performance parameters were investigated through \apsq simulations, where the collected charge between two strips in an ELAD detector was modelled as a function of MIP incidence position.
As seen in Figure~\ref{fig:elad_resolution}, the theoretical optimum is a linear charge sharing between the two strips, with maximal sharing when the particle is incident at the center between both strips.
The ELAD simulations are much closer to this calculated optimum than the standard case, thus it can improve upon the binary limit of $strip\_distance/\sqrt{12}$.
In another \apsq simulation, the ELAD detector was found to have a factor of two improvement from the binary positional resolution compared to a standard planar sensor for a pixel pitch of \SI{55}{\um}.
Further simulation studies using both softwares are currently ongoing.

\textbf{Assembly production :} Due to the multiple layers needed of ion implantation and epitaxial growth, the manufacture of ELADs is challenging.
The first set of ELAD assemblies are in production to demonstrate the performance capabilities of this concept.

%% file: chapters/monolithic.tex
\section{Monolithic technologies for the tracking detector}
\label{ch:monolithic}
\subsection{CLICTD}
\textbf{Detector design :} The monolithic CLIC Tracking Detector (CLICTD) is produced using a \SI{180}{\nm} CMOS imaging process with a HR expitaxial sensor layer~\cite{clictd_proceedings}.
It features a 128$\times$16 matrix with pixels of pitch \SI{300}{\um}$\times$\SI{30}{\um}.
Each pixel is further segmented into eight subpixels of size \SI{37.5}{\um}$\times$\SI{30}{\um}, each with an analogue front-end but a shared digital front-end.
This design aims to decrease the charge collection time while mainting a low amount of digital circuitry per pixel.
Three pixel modes are implemented: simultaneous 8-bit integrated ToT and 5-bit ToA; 13-bit ToA; and event counts.
These measurements are performed in the on-channel digital logic for the combined output from the subpixels by means of an "OR" gate.
The data readout uses a shutter-based scheme.

\begin{figure}[tb]
\centering
\begin{minipage}{.48\textwidth}
    \centering
    \includegraphics[width=\linewidth]{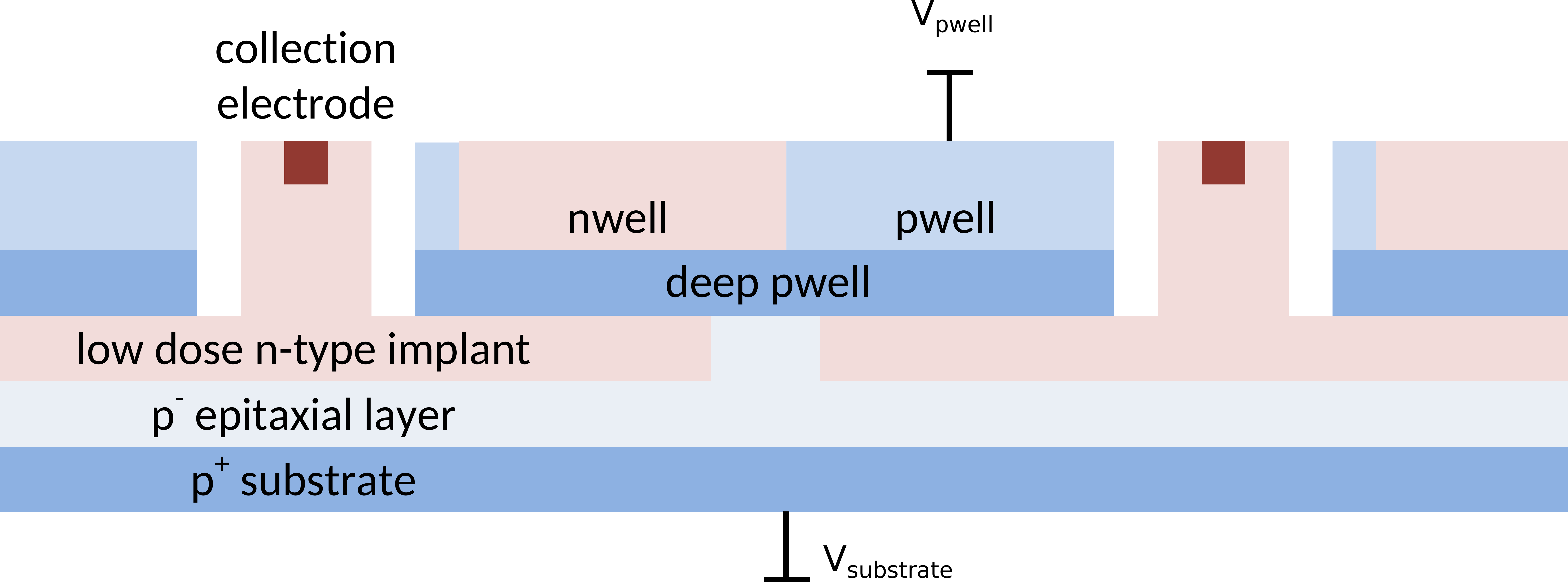}
    \caption{Diagram of the CLICTD detector with a continous N layer design~\cite{clictd_proceedings}.}
    \label{fig:clictd_1}
\end{minipage}%
\hfill
\begin{minipage}{.48\textwidth}
    \centering
    \includegraphics[width=\linewidth]{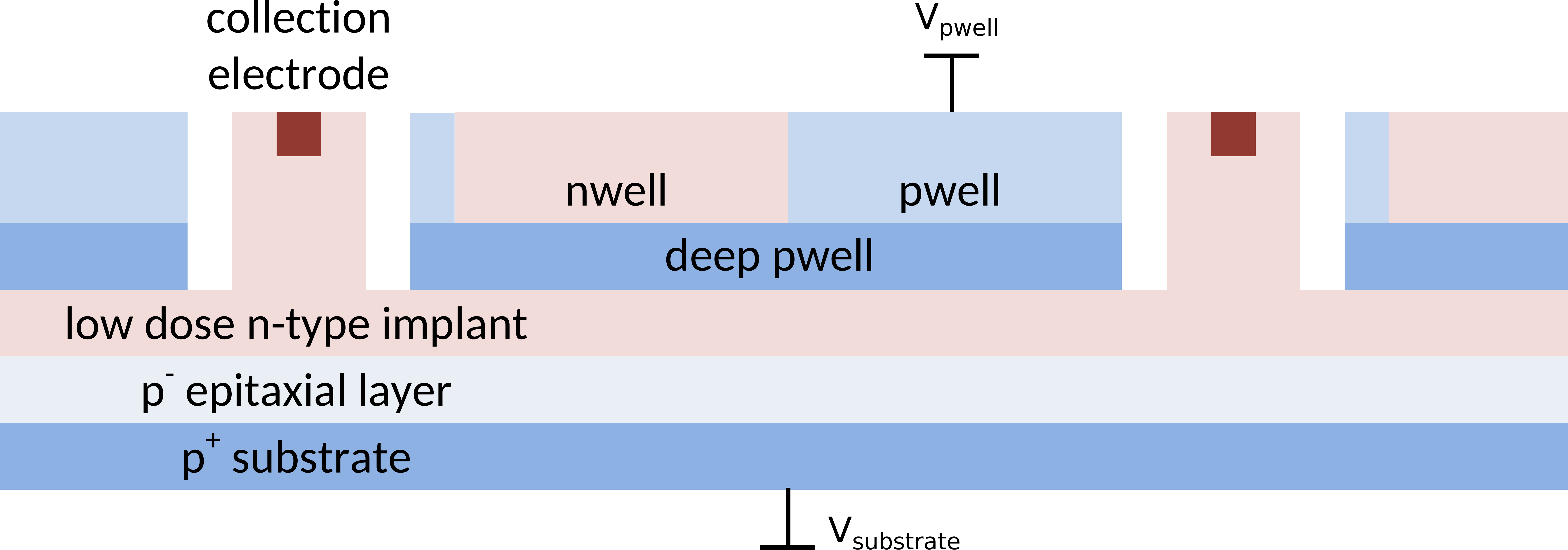}
    \caption{Diagram of the CLICTD detector with an additional N layer gap design~\cite{clictd_proceedings}.~}
    \label{fig:clictd_2}
\end{minipage}  
\end{figure}

Figure~\ref{fig:clictd_1} presents the process cross-section of two pixels in CLICTD.
Charge is collected by a small collection electrode situated inside a low-dose N-type implant, itself placed above a P-type expitaxial layer~\cite{clictd_proceedings}.
To ensure full lateral depletion, a HR epitaxial layer is used.
A modified design has introduced a gap in the N layer to reduce the charge collection time further, this design is shown in Figure~\ref{fig:clictd_2}.
The small electrode capacitance in the order of fF results in a low expected detection threshold of approximately \SI{180}{electrons}, low power consumption, and a large signal-to-noise ratio.

\textbf{Simulation results :} Due to the complex, non-linear nature of the electric field in this device, the design of CLICTD was optimised through detailed TCAD simulations~\cite{detector_tech_report}.
Currently, Monte Carlo (MC) simulations using \apsq are being carried out to better understand the response of the samples in laboratory and test beam measurements.

\begin{figure}[tb]
\centering
\begin{minipage}{.48\textwidth}
    \includegraphics[width=1.05\linewidth]{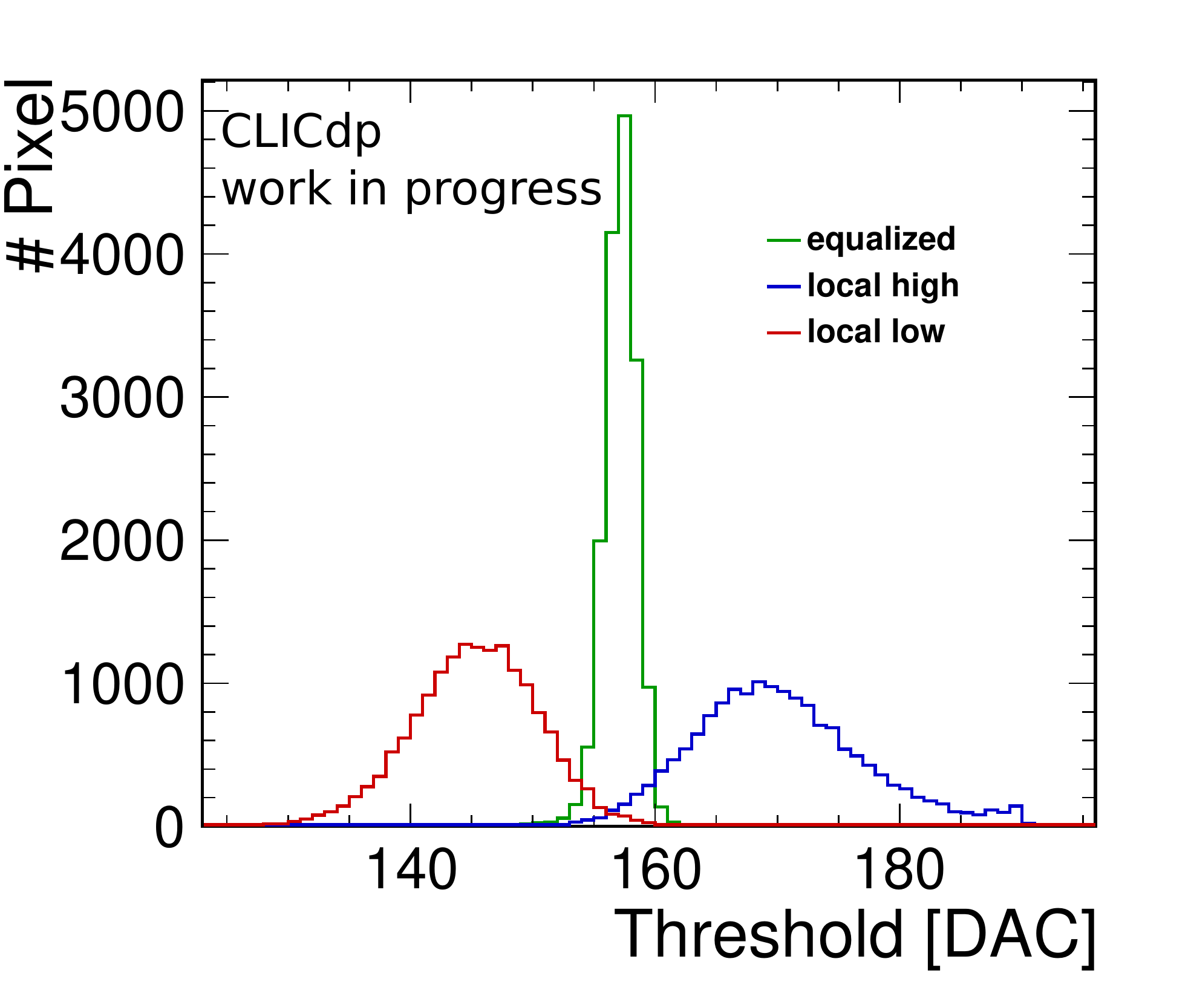}
    \caption{Equalisation histograms for CLICTD sample B1.
    Data was taken at maximum and minimum local threshold settings to determine the equalised threshold adjustment for each pixel.}
    \label{fig:clictd_equalisation}
\end{minipage}%
\hfill
\begin{minipage}{.48\textwidth}
    \centering
    \includegraphics[width=1.05\linewidth]{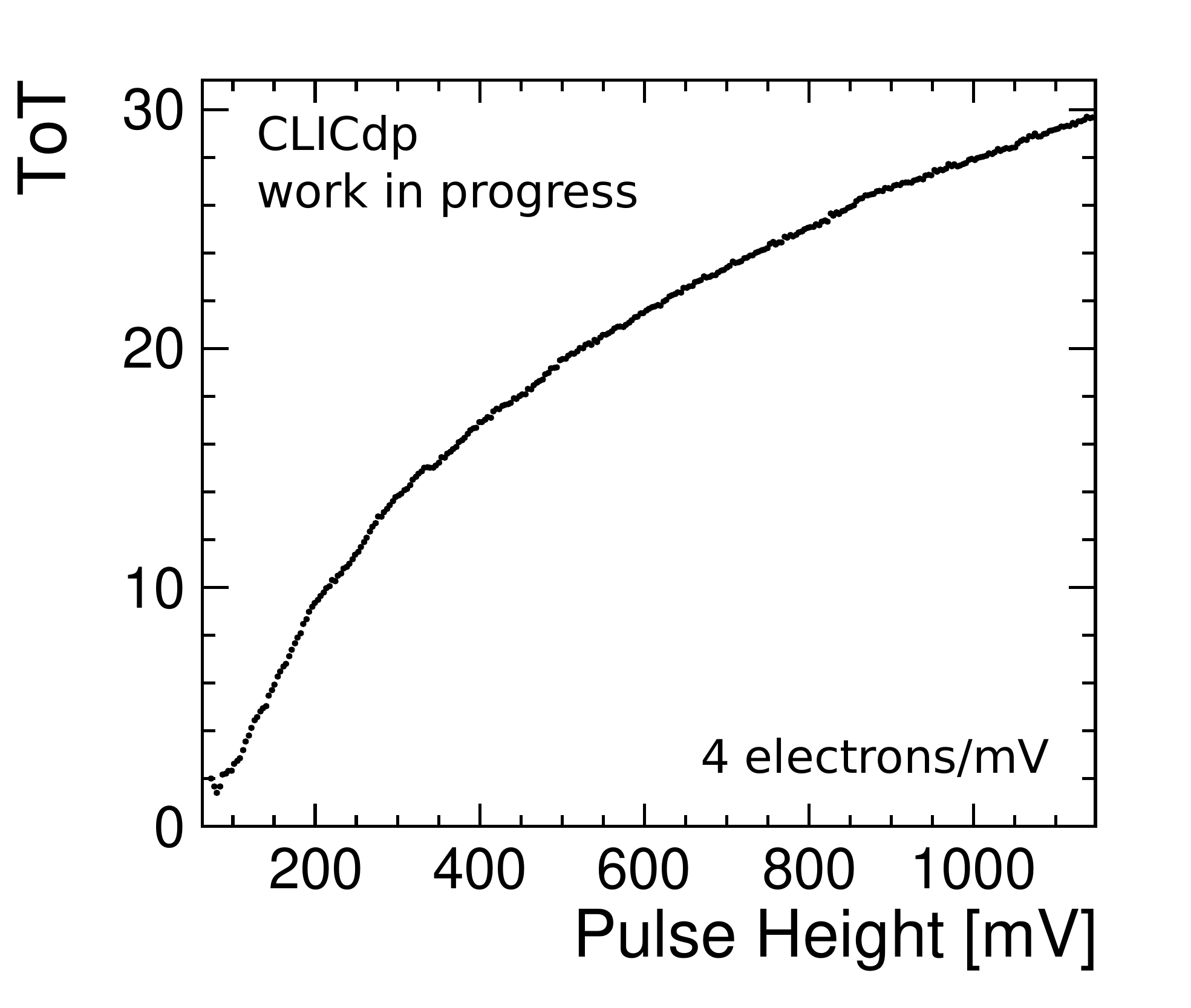}
    \caption{Histogram of ToT value for pixel 0,0 of CLICTD sample A1 when test pulses of different magnitudes injected charge into the pixel.}
    \label{fig:clictd_tot}
\end{minipage}  
\end{figure}

\textbf{Laboratory results :} CLICTD samples have been produced using both process variants.
The I-V characteristics of these assemblies were measured for multiple P well bias voltages.
It was found that the substrate in the continuous deep N-type implant design can be biased to higher magnitudes due to a better isolation between the substrate and the P wells~\cite{clictd_proceedings}.
After equalisation of the pixel matrix, a threshold dispersion of approximately \SI{16}{electrons} was extracted for sample B1, see Figure~\ref{fig:clictd_equalisation}.
This assembly was produced with the additional N layer gap design, has an average noise of approximately \SI{18}{electrons}, and was operated at a p-well/substrate bias voltage of -3/\SI{-3}{\V}.
The ToT response of a single pixel from test pulsing measurements of CLICTD sample A1 is presented in Figure~\ref{fig:clictd_tot}.
Assembly A1 has the continuous N layer design, has an average noise of approximately \SI{20}{electrons}, and a threshold dispersion of \SI{16}{electrons}.
It was operated at a threshold of approximately \SI{330}{electrons} and a p-well/substrate bias voltage of -3/\SI{-6}{\V}.
CLICTD chips have been exposed to fluorescent X-rays produced from multiple target materials to determine the conversion factor between threshold DAC value and the threshold level in electrons, which was found to be \SI{15}{electrons/DAC} and \SI{14}{electrons/DAC} for assemblies A1 and B1 respectively~\cite{clictd_proceedings}.

\subsection{ATLASpix}
Several monolithic HV-CMOS sensors have been developed for the High-Luminosity upgrade of the ATLAS pixel detector outer layer.
The ATLASpix\_Simple sensor is interesting to CLIC as it targets similar requirements to those of the CLIC tracking detector.
Assemblies with resistivities from \SI{20}{\ohm\cm} to \SI{200}{\ohm\cm}, and total thicknesses of \SI{62}{\um} and \SI{100}{\um}, have been studied.

\textbf{Detector design :} This HV monolithic active pixel sensors (HV-MAPS) was produced using a commerical \SI{180}{\nm} HV-CMOS process.
Elongated pixels of pitch \SI{130}{\um}$\times$\SI{40}{\um} are arranged in a matrix of 25 columns and 400 rows.
ATLASpix\_Simple has a fully integrated data-driven readout and a fast charge collection using drift.
Each pixel can record a 10-bit ToA and 6-bit ToT.

\begin{figure}[tb]
\begin{minipage}[t]{0.48\textwidth} 
    \centering
    \includegraphics[width=\linewidth]{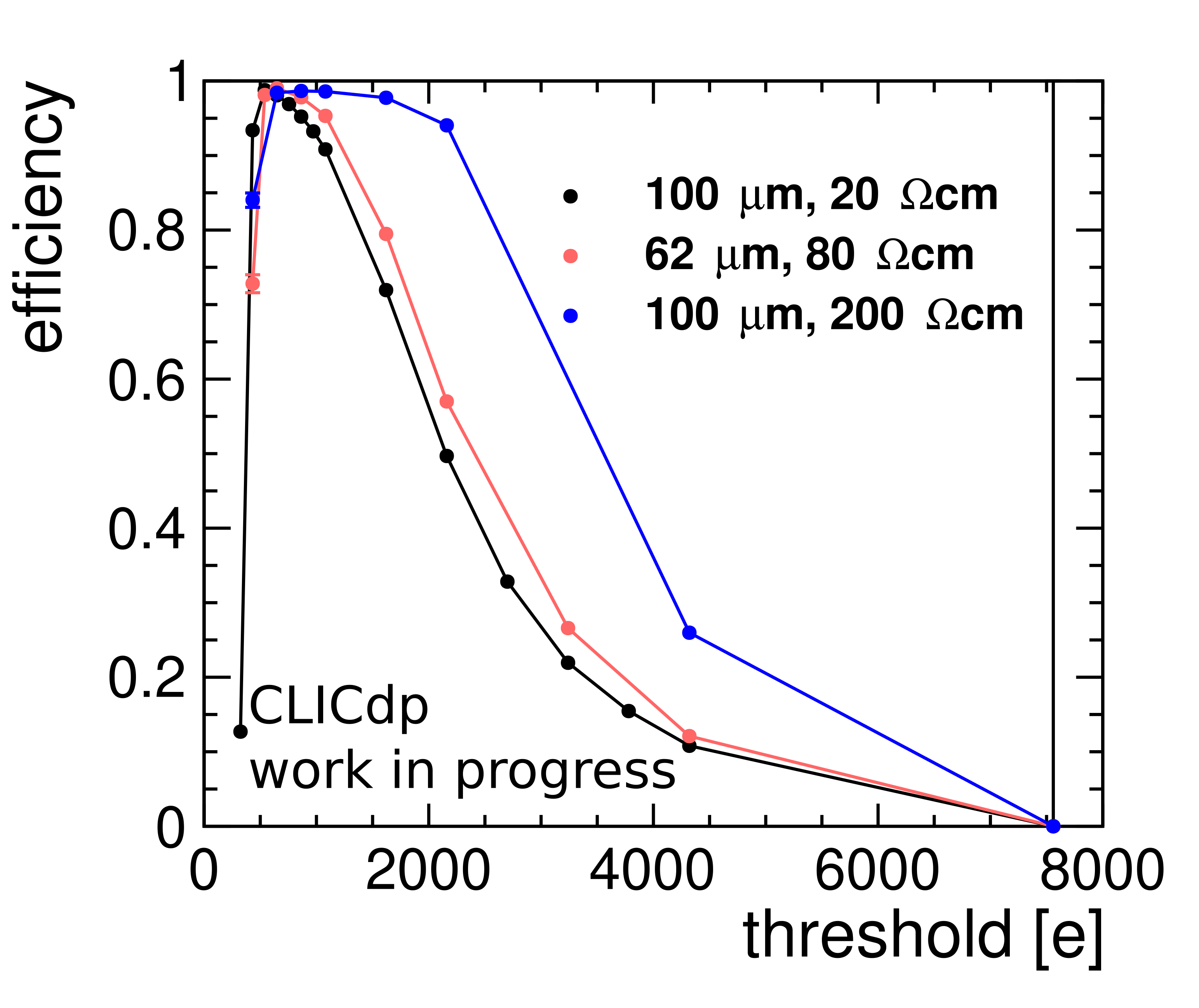}
    \caption{Efficiency as a function of threshold for three ATLASpix\_Simple detectors.}
    \label{fig:atlaspix_eff}
 \end{minipage}\hfill
 \begin{minipage}[t]{0.48\textwidth}
    \centering
    \includegraphics[width=\linewidth]{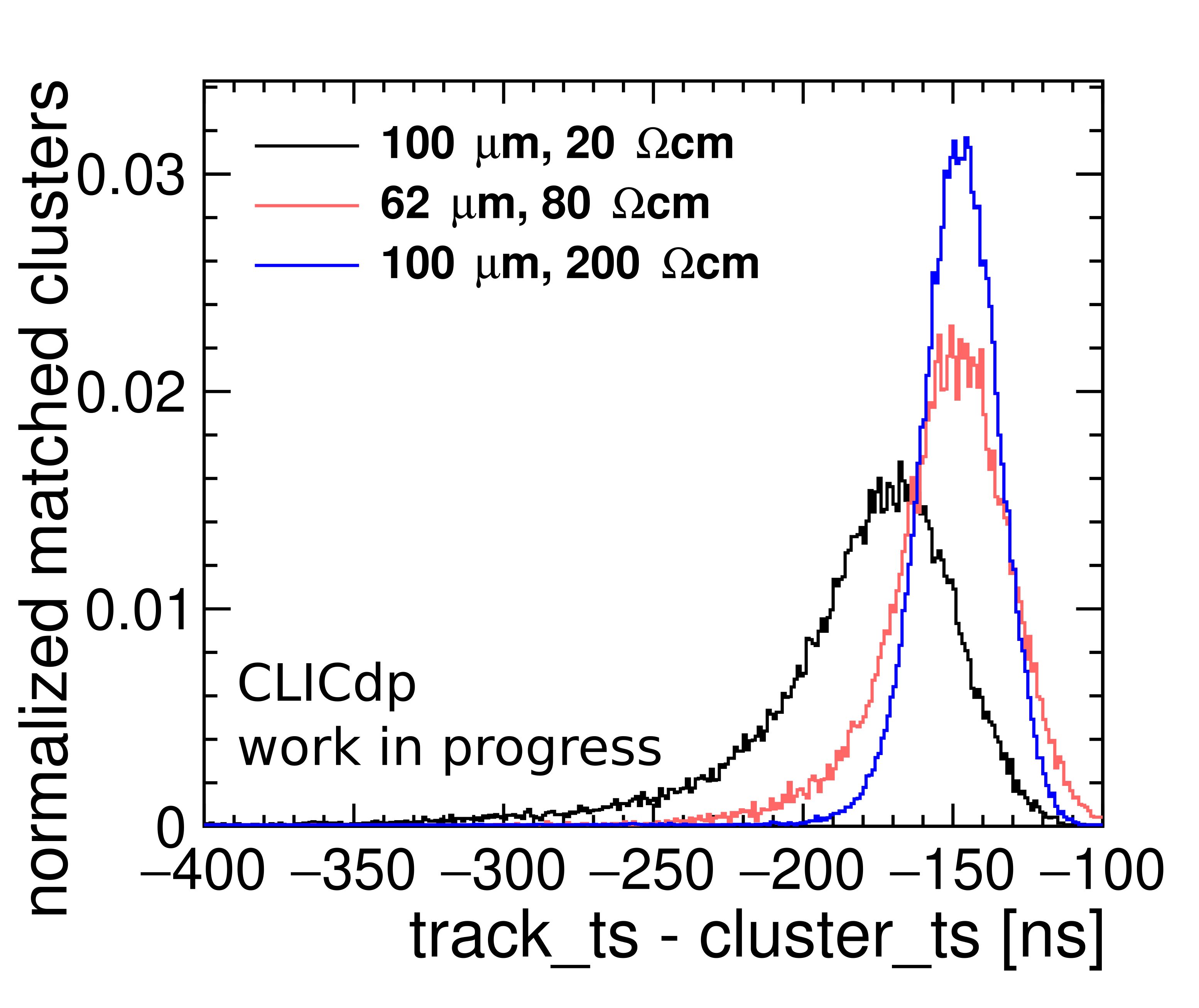}
    \caption{Time residual distributions for three ATLASpix\_Simple detectors.}
    \label{fig:atlaspix_timing}
 \end{minipage}
\end{figure}

\textbf{Test-beam results :} Samples of three different resistivities and two different thicknesses were tested at the DESY test-beam facility~\cite{DESYpaper} to evalue their performance.
Each device was nominally operated at a bias voltage of \SI{-50}{\V} and a threshold of approximately \SI{650}{electrons}.
The data curves in Figure~\ref{fig:atlaspix_eff} show the distributions of efficiency vs. threshold level for three samples.
Efficiencies of 98-99\% were achieved for the nominal operational conditions.
Compared to the other resistivities, the \SI{200}{\ohm\centi\meter} assembly has a larger plataeu region before the efficiency decreases for higher thresholds due to the larger signal.
The time residuals between the track and cluster timestamps for the same three samples are shown in Figure~\ref{fig:atlaspix_timing}.
The \SI{20}{\ohm\centi\meter} detector has a slower timing response and is subject to a larger timewalk effect when compared to the other detectors, as expected due to the lower amount of charge collected by the smaller depleted volume.

%% file: chapters/toolsdeveloped.tex
\section{Characterisation tools developed}
\label{ch:toolsdeveloped}
Multiple tools have been developed within  CLICdp to carry out detailed pixel characterisation studies, such as laboratory testing, MC device simulations, and test-beam data reconstruction.

\textbf{Caribou:} A flexible readout system for pixel detectors has been developed and is used for laboratory testing of novel devices~\cite{caribou}.
The Caribou Data AQuisition (DAQ) system hardware consists of a custom chip board that is specific to the type of device, connected to the universal periphery Control and Readout (CaR) board, providing the hardware environment for various target ASICs.
The CaR board is then connected to a System on Chip (SoC) that contains both a CPU for data aquisition, running a full linux system, and a FPGA fabric for detector control and data handling.
It is noted that no dedicated PC is required, only an ethernet link is needed and the user can access the system via a Secure SHell (SSH).
The use of a versatile and well tested DAQ, rather than developing a device specific system each time, significantly reduces the time needed for protoype developement and testing.

\textbf{Allpix$^2$:} To better understand the behaviour of complex detectors, a modular and generic framework for Monte-Carlo simulations has been developed~\cite{allpix-squared}.
This comprehensive software can model detector response from incident ionising particles through to digitisation in the readout system. 
Performance parameters of detectors can be precisely modelled due to the ability to import TCAD~\cite{synopsys-tcad} simulated electric fields and combine them with simplified models of charge transportation.
The modular framework design allows complex systems of multiple detectors to be simulated in a flexible way.
Detailed information on \apsq can be found on the software repository~\cite{allpix-squared-gitlab} or on the \apsq website~\cite{allpix-squared-website}.

\textbf{Corryvreckan:} A modular, lightweight reconstruction software framework has been created to analyse test-beam data~\cite{corryvreckan-gitlab}.
\corry aims to be flexible, fast, and have high levels of documentation for ease of use.
Many telescope configurations can be described as the software can easily combine data streams of different triggered, data-driven, and frame-based readout schemes.
This event building method also means that the efficiency of each detector can be calculated accurately in such a complex set-up.
One key feature of \corry is its utilisation of timing information provided by sensors.
Track building, clustering, and DUT-track association algorithms can all use timing information to decrease combinatorics and recover clusters accurately in high-rate environments.
A v1.0 release is available, as detailed on the \corry website~\cite{corryvreckan-webpage}, and the v1.0 users manual can be found as a published note~\cite{corry_manual_paper}.

%% file: chapters/conclusion.tex
\section{Conclusion and outlook}
\label{ch:conclusion}
A diverse R\&D programme is underway in the CLICdp collaboration on novel hybrid and monolithic technologies to fulfil the vertex and tracking detector requirements respectively.
High quality CLICpix2 hybrid detectors have been produced with a interconnect yields of up to 97.9\%.
The pixel categorisation method employed has allowed the data from multiple laboratory tests to be combined, providing more detail on the exact response of individual pixels.
An optimal operational voltage of \SI{-25}{\V} and an efficiency of 98.1\% has been calculated for one sample from test-beam data analysis.
The ELAD sensor aims to increase charge sharing between pixels with a fixed pixel size and sensor thickness.
In-depth TCAD and MC simulations have predicted that a near linear charge sharing can be achieved, improving the expected positional resolution of the device by a factor of two compared to standard planar sensors.
CLICTD monolithic assemblies have been produced and show promising laboratory results.
Further testing, including test-beam data reconstruction and simulations, is ongoing to better understand the performance parameters of the detector.
Test-beam data has been taken for monolithic ATLASpix\_Simple assemblies of different resistivities and thicknesses.
Efficiencies of 98-99\% were achieved and the timing behaviour of the samples has been analysed in detail.
Through these investigations a set of flexible tools for pixel detector characterisation have been developed for DAQ, MC simulation, and test-beam data reconstruction.

%% file: chapters/acknowledgements.tex
\section*{Acknowledgements}
\textit{The measurements leading to these results have been performed at the Test Beam Facility at DESY Hamburg (Germany), a member of the Helmholtz Association (HGF).
This work was supported by the European Union's Horizon 2020 Research and Innovation programme under Grant Agreement No. 654168 (AIDA-2020).}

%% file: ms.bbl
\providecommand{\href}[2]{#2}\begingroup\raggedright\begin{thebibliography}{10}

\bibitem{CLIC-summary-report-2018}
{\scshape CLIC} collaboration, {P.N. Burrows et al. (editors)}, \emph{{The
  Compact Linear Collider (CLIC) - 2018 Summary Report}},
  \href{http://dx.doi.org/10.23731/CYRM-2018-002}{\emph{CERN Yellow Rep.
  Monogr.} {\bfseries 1802} (Dec, 2018) 01--98},
  [\href{https://arxiv.org/abs/1812.06018}{{\ttfamily 1812.06018}}].

\bibitem{detector_tech_report}
{\scshape CLIC} collaboration, {D. Dannheim et al. (editors)}, \emph{{Detector
  Technologies for CLIC}},
  \href{http://dx.doi.org/10.23731/CYRM-2019-001}{\emph{CERN Yellow Rep.
  Monogr.} (May, 2019) }, [\href{https://arxiv.org/abs/1905.02520}{{\ttfamily
  1905.02520}}].

\bibitem{CLICdet}
N.~Alipour~Tehrani et~al., \emph{{CLICdet: The post-CDR CLIC detector model}},
  CLICdp-Note-2017-001, 28 Feb, {2018 (revised 05 April, 2019)}.

\bibitem{powerpusling_tpx3}
E.~Perez~Codina, \emph{{Timepix3 performance in power pulsing operation}},  in
  \emph{{Proceedings of 27th International Workshop on Vertex Detectors
  (VERTEX): Chennai, India, Oct 21-26, 2018}}, vol.~VERTEX2018, p.~59, 2019,
  DOI:10.22323/1.348.0059.

\bibitem{powerpusling_cpx2}
S.~Ruiz~Daza, ``{Power pulsing with CLICpix2}.'' CERN-STUDENTS-Note-2019-212,
  \url{https://cds.cern.ch/record/2690112}, 19 Sep, 2019.

\bibitem{cpx2_manual}
E.~Santin, P.~Valerio and A.~Fiergolski, ``{CLICpix2 User's Manual v1.0}.''
  \url{https://edms.cern.ch/document/1800546/1}, 24 April, 2017.

\bibitem{IZM}
``{Fraunhofer Institute for Reliability and Microintegration (IZM) website}.''
  \url{https://www.izm.fraunhofer.de/en.html}, Accessed: Dec, 2019.

\bibitem{FBK}
``{Fondation Bruno Kessler, Center for Materials and Microsystems (FBK-CMM)
  website}.'' \url{https://cmm.fbk.eu/en/}, Accessed: Dec, 2019.

\bibitem{Advacam}
``{ADVACAM website}.'' \url{http://www.advacam.com}, Accessed: Dec, 2019.

\bibitem{EUDET_telescope_paper}
H.~Jansen et~al., \emph{{Performance of the EUDET-type beam telescopes}},
  \href{http://dx.doi.org/10.1140/epjti/s40485-016-0033-2}{\emph{{EPJ
  Techniques and Instrumentation}} {\bfseries 3} (10, 2016) 07,
  DOI:10.1140/epjti/s40485--016--0033--2}.

\bibitem{DESYpaper}
R.~Diener et~al., \emph{{The DESY II test beam facility}},
  \href{http://dx.doi.org/{10.1016/j.nima.2018.11.133}}{\emph{NIMA} {\bfseries
  922} (April, 2019) 265--286, DOI:10.1016/j.nima.2018.11.133}.

\bibitem{elad}
A.~Velyka and H.~Jansen, \emph{{Enhanced Lateral Drift Sensors: Concept and
  Development}},  in \emph{{Proceedings of International Conference on
  Technology and Instrumentation in Particle Physics 2017 (TIPP): Beijing,
  China, May 22-26, 2017}}, vol.~213, pp.~380--384, {2018,
  DOI:10.1007/978-981-13-1316-5$\_$71}.

\bibitem{synopsys-tcad}
``{Synopsys TCAD website}.'' \url{https://www.synopsys.com/silicon/tcad.html},
  Accessed: Dec, 2019.

\bibitem{clictd_proceedings}
I.~Kremastiotis et~al., \emph{{CLICTD: A monolithic HR-CMOS sensor chip for the
  CLIC silicon tracker}},  No.~CLICdp-Conf-2019-009,
  \url{http://cds.cern.ch/record/2693182}, 11 Oct, 2019.

\bibitem{caribou}
T.~Vanat, \emph{{Caribou - A versatile data acquisition system}},
  No.~CLICdp-Conf-2019-012, \url{http://cds.cern.ch/record/2703500}, 05 Dec,
  2019.

\bibitem{allpix-squared}
S.~Spannagel et~al., \emph{{Allpix$^2$: A modular simulation framework for
  silicon detectors}},
  \href{http://dx.doi.org/10.1016/j.nima.2018.06.020}{\emph{NIMA} {\bfseries
  A901} (2018) 164 -- 172, DOI:10.1016/j.nima.2018.06.020}.

\bibitem{allpix-squared-gitlab}
``{\apsq software repository}.''
  \url{https://gitlab.cern.ch/allpix-squared/allpix-squared}, Accessed: Dec,
  2019.

\bibitem{allpix-squared-website}
``{Allpix$^2$ website}.'' \url{http://project-allpix-squared.web.cern.ch/},
  Accessed: Dec, 2019.

\bibitem{corryvreckan-gitlab}
``{Corryvreckan software repository}.''
  \url{https://gitlab.cern.ch/corryvreckan/corryvreckan}, Accessed: Dec, 2019.

\bibitem{corryvreckan-webpage}
``{Corryvreckan website}.'' \url{http://project-corryvreckan.web.cern.ch},
  Accessed: Dec, 2019.

\bibitem{corry_manual_paper}
J.~Kroeger, S.~Spannagel and M.~Williams, ``{User Manual for the Corryvreckan
  Test Beam Data Reconstruction Framework, Version 1.0}.''
  \url{https://cds.cern.ch/record/2703012}, 28 Nov, 2019.

\end{thebibliography}\endgroup
